\documentclass[aps,preprint,nobibnotes,nofootinbib]{revtex4-1}
\usepackage{amssymb,amsfonts,amsmath,amsthm}
\usepackage[all,arc]{xy}
\usepackage{enumerate}
\usepackage{mathrsfs}
\usepackage{graphicx}

\begin{document}

\title{A Note on the Coordinate Freedom in Describing the Motion of Particles in General Relativity}

\author{Samuel E. Gralla}
\author{Robert M. Wald}
\affiliation{\it Enrico Fermi Institute and Department of Physics \\ \it University of Chicago
\\ \it 5640 S.~Ellis Avenue, Chicago, IL~60637, USA}

\begin{abstract}
Our previous work developed a framework for treating the motion of a
small body in general relativity, based on a one-parameter-family of
solutions to Einstein's equation. Here we give an analysis of the 
coordinate freedom allowed within this framework, as is needed to determine the 
form of the equations of motion when they are expressed in general gauges.

\end{abstract}

\maketitle

In \cite{gralla-wald}, we analyzed particle motion in
general relativity by considering a one-parameter family of metrics,
$g_{ab}(\lambda)$, corresponding to a body that shrinks to zero size
and mass in an asymptotically self-similar manner. To express this
precisely, we introduce coordinates $(t,x^i)$ and denote the metric
components in these coordinates as $g_{\mu \nu}(\lambda,x)$. We write
$r=\sqrt{\delta_{ij}x^i x^j}$, and we replace the variables
$(\lambda,t,x^i)$ by $\alpha=r$, $\beta=\lambda/r$, $n^i=x^i/r$
(denoting a direction on the sphere) and $t$. The main requirement of
\cite{gralla-wald} on the one-parameter family of metrics is that the
metric components $g_{\mu \nu}$ be smooth\footnote{This requirement
  was stated in \cite{gralla-wald} as that of joint smoothness
  $(\alpha,\beta)$ at fixed $(n^i,t)$, but joint smoothness in
  $(\alpha,\beta,n^i,t)$ is actually needed.} in the variables
$(\alpha,\beta,n^i,t)$, including at $\alpha = \beta = 0$; see  \cite{gralla-wald} for a full discussion
of why this is an appropriate requirement to describe a one parameter family of bodies that shrink down to $x^i=0$ in the limit as $\lambda \to 0$.  We also
require that at $\lambda=0$, $g_{\mu \nu}$ is smooth in $(t,x^i)$ and
that the worldline defined by $x^i = 0$ is timelike.  

In \cite{gralla-wald}, we showed that the worldline $x^i = 0$ must be
a geodesic of $g_{ab}(\lambda=0)$, and we derived the corrections to
the motion of the center of mass to first order in $\lambda$ in the Lorenz gauge. It is of
interest to describe these corrections to the motion in other gauges,
and, for this reason, it is of interest to determine the class of
coordinate transformations that preserve our above requirements. In new coordinates $x'^{\mu'} (\lambda, x^\mu)$, the metric components take the form
\begin{equation}
g'_{\mu' \nu'}(\lambda, x') = g_{\mu \nu}(\lambda, x) \frac{\partial x^\mu}{\partial x'^{\mu'}} \frac{\partial x^\nu}{\partial x'^{\nu'}} \, .
\end{equation}
The allowed coordinate transformations are those for which the new metric components
$g'_{\mu' \nu'}$ are (1) smooth in $(\alpha',\beta',n'^i,t')$, and (2) smooth in $(t', x'^i)$ at $\lambda = 0$.
In \cite{gralla-wald}, it was asserted, without any attempt at justification,
that the allowed coordinate transformations are those  for which
$x'^{\mu'}(\lambda, x^\nu)$ is jointly smooth in $(\lambda, x^\nu)$
for all $r > C \lambda$ for some constant $C$
and for which the Jacobian 
$\partial x'^{\mu'}/\partial x^{\mu}$ is smooth in $(\alpha,\beta,n^i,t)$. The purpose of this note is to give a more
careful and complete analysis of this issue. We will thereby provide justification for (but not a complete proof of) the claim of \cite{gralla-wald}---subject to a caveat explained below---and we will provide the form of the allowed transformations in much more explicit detail. Specifically, we shall argue that---with the exception of some very special cases, such as when $g_{ab}(\lambda)$ is flat for all $\lambda$, where a limited class of additional transformations are possible---the general allowed coordinate transformations are those of the form
\begin{equation}
x'^\mu = f_1^\mu(x^\nu) + \lambda f_2^\mu(\alpha,\beta,n^i,t) \, ,
\label{allowed2}
\end{equation}
where $f_1$ is a diffeomorphism that leaves the origin ($x^i = 0$) fixed and $f_2$ is a smooth function of its arguments.
 
Since at $\lambda=0$ the metric components in the new coordinates must be smooth in the new coordinates, the coordinate transformation at $\lambda = 0$ must be a diffeomorphism. By applying the inverse of this diffeomorphism at all $\lambda$, we may assume without loss of generality 
that $x'^{\mu}(\lambda=0) = x^\mu$. Now, for $r> C \lambda$ (i.e., for $\alpha > 0$ and $\beta \leq 1/C$)
the variables $(\lambda, x^\mu)$ are smoothly related to $(\alpha,\beta,n^i,t)$, and similarly for the primed variables. It follows that the allowed transformations must take the form
\begin{equation}
x'^\mu = x^\mu + \lambda \mathcal{F}^\mu(\lambda,t,x^i),
\end{equation}
where ${\mathcal F}^\mu$ is smooth in its arguments for $r> C \lambda$ (but 
$\mathcal{F}^\mu$ need not be smoothly extendable to $r=\lambda=0$). Equivalently, since $\lambda = \alpha \beta$ and $x^i=r n^i$ we may write
\begin{equation}\label{mainform}
x'^\mu = x^\mu + \alpha \beta F^\mu(\alpha,\beta,n^i,t), 
\end{equation}
where $F^\mu$ is smooth in its arguments for $\alpha > 0$ and $\beta \leq 1/C$. We now investigate the consequences of imposing the condition, proposed in \cite{gralla-wald}, that the Jacobian 
$\partial x'^{\mu'}/\partial x^{\mu}$ be smooth in $(\alpha,\beta,n^i,t)$ (including at $\alpha=\beta=0$).  We will prove that a coordinate transformation of the form \eqref{mainform} for which $\partial x'^{\mu'}/\partial x^{\mu}$ is smooth in $(\alpha,\beta,n^i,t)$ must take the form of equation \eqref{coords}.

In terms of equation \eqref{mainform} the Jacobian is computed to be
\begin{align}
\frac{\partial x'^\mu}{\partial t} & = \alpha \beta \frac{\partial F^\mu}{\partial t} \label{jact} \\
\frac{\partial x'^\mu}{\partial x^i} & = \delta^\mu_{\ i} + \beta n_i \left( \alpha \frac{\partial F^\mu}{\partial \alpha} - \beta \frac{\partial F^\mu}{\partial \beta} \right) + \alpha \beta \left( \delta^j_{\ i} - n^j n_i \right) \frac{\partial F^\mu}{\partial n^j}. \label{jacx}
\end{align}  
We begin with equation \eqref{jact}. Since the right side must smoothly extend to $\alpha=\beta=0$, we have
\begin{equation}\label{uset}
\frac{\partial F^\mu}{\partial t} = \frac{S^\mu(\alpha,\beta,n^i,t)}{\alpha \beta},
\end{equation}
where $S^\mu$ is a smooth function of its arguments, including at $\alpha=\beta=0$. 
In our calculations here and below, we consistently use the letter ``$S$'' (possibly with subscripts or other modifiers) to denote a function known to be smooth in its arguments including at $\alpha=\beta=0$, and we consistently use the letter ``$F$'' (possibly with subscripts or other modifiers) to denote a function known to be smooth in its arguments for $\alpha > 0$ and $\beta \leq 1/C$, but not necessarily at $\alpha = \beta = 0$. 
Since $S^\mu$ is smooth, we may write 
\begin{equation}
S^\mu(\alpha,\beta,n^i,t) = S_0^\mu(\alpha,n^i,t) + \beta S_1^\mu(\alpha,\beta,n^i,t),
\end{equation}
where $S_0^\mu$ and $S_1^\mu$ are smooth in their arguments, whence equation \eqref{uset} becomes
\begin{equation}
\frac{\partial F^\mu}{\partial t} = \frac{S_0^\mu(\alpha,n^i,t)}{\alpha \beta}  + \frac{S_1^\mu(\alpha,\beta,n^i,t)}{\alpha}.
\end{equation}
However, since $F^\mu$ is smooth in $\beta$ at fixed $\alpha>0$, we must have $S^\mu_0=0$, so we in fact have
\begin{equation}\label{usetnob}
\frac{\partial F^\mu}{\partial t} = \frac{S_1^\mu(\alpha,\beta,n^i,t)}{\alpha},
\end{equation}
We may now integrate \eqref{usetnob} to learn
\begin{equation}\label{aftert}
F^\mu = \frac{1}{\alpha} \tilde{S}_1^\mu(\alpha,\beta,n^i,t) + F_0^\mu (\alpha,\beta,n^i) \, .
\end{equation}

Now consider the spatial part of the Jacobian, equation \eqref{jacx}.  The parts parallel and perpendicular to $n^i$ (as measured by $\delta_{ij}$) must separately be smooth.  Beginning with the perpendicular part, we obtain in parallel with the derivation of \eqref{usetnob} that
\begin{equation}\label{usexp}
\left( \delta^j_{\ i} - n^j n_i \right)\frac{\partial F^\mu}{\partial n^j} = \frac{S^\mu_{i}(\alpha,\beta,n^i,t)}{\alpha},
\end{equation}
for some smooth $S^\mu_i$.  Substituting from equation \eqref{aftert} on the left side, we obtain
\begin{equation}
\left( \delta^j_{\ i} - n^j n_i \right) \frac{\partial}{\partial n^j} F_0^\mu(\alpha,\beta,n^i) = \frac{\tilde{S}^\mu_i(\alpha,\beta,n^i)}{\alpha}
\end{equation}
for some smooth $\tilde{S}^\mu_i$. Integrating this equation, we find that $F_0^\mu$ takes the form
\begin{equation}
F_0^\mu = F_{00}^\mu(\alpha,\beta) + \frac{\bar{S}^\mu(\alpha,\beta,n^i)}{\alpha} \, .
\end{equation}
Then we can substitute into equation \eqref{aftert} to find
\begin{equation}\label{afterxp}
F^\mu = \frac{1}{\alpha} S'^\mu(\alpha,\beta,n^i,t) + F_{00}^\mu (\alpha,\beta) \, .
\end{equation}
It will be convenient to write
\begin{equation}
S'^\mu(\alpha,\beta,n^i,t) = S'^\mu_0(\beta,n^i,t) + \alpha S'^\mu_1(\alpha,\beta,n^i,t) \, ,
\end{equation}
whence we obtain
\begin{equation}\label{afterxp2}
F^\mu = \frac{1}{\alpha} S'^\mu_0(\beta,n^i,t) + S'^\mu_1(\alpha,\beta,n^i,t) + F_{00}^\mu (\alpha,\beta) \, .
\end{equation}
We may assume without loss of generality that $S'^\mu_0(\beta,n^i,t)$ has nontrivial dependence on $n^i$ and/or
$t$, since, otherwise, we could absorb $S'^\mu_0/\alpha$ into $F_{00}^\mu$.

Finally consider the part parallel to $n^i$ of equation \eqref{jacx},
\begin{equation}\label{usexl}
\beta \left( \alpha \frac{\partial F^\mu}{\partial \alpha} - \beta \frac{\partial F^\mu}{\partial \beta} \right) = \hat{S}^\mu(\alpha,\beta,n^i,t).
\end{equation}
We substitute \eqref{afterxp2} in equation \eqref{usexl} to find
\begin{equation}\label{oinkpig}
\beta \left[ \alpha \frac{\partial F^\mu_{00}}{\partial \alpha} - \beta \frac{\partial F_{00}^\mu}{\partial \beta} - \frac{1}{\alpha} \left( {S'}_0^\mu + \beta \frac{\partial {S'}_0^\mu}{\partial \beta} \right) \right] = \hat{S}^\mu(\alpha,\beta,n^i,t),
\end{equation}
where we have absorbed the smooth terms arising from ${S'}_1^\mu$ into $\hat{S}^\mu$.  Since $F_{00}^\mu$ is independent of $(n^i,t)$ and ${S'}_0^\mu$ has nontrivial dependence on these variables, equation \eqref{oinkpig} can only be satisfied if
\begin{equation}
\frac{\beta}{\alpha} \left( S'^\mu_0 + \beta \frac{\partial S'^\mu_0}{\partial \beta} \right) = 0,
\end{equation}
which can be integrated to find $S'^\mu_0 = \hat{F}^\mu(n^i,t)/\beta$, implying that in fact $S'^\mu_0=0$ (since $S'^\mu_0$ is smooth).  Thus, we obtain
\begin{equation}\label{afterxl}
F^\mu = \tilde{S}^\mu(\alpha,\beta,n^i,t) + F_{00}^\mu (\alpha,\beta)
\end{equation}
and
\begin{equation}
\alpha \beta \left( \frac{\partial F_{00}^\mu}{\partial \alpha} - \frac{\beta}{\alpha} \frac{\partial F_{00}^\mu}{\partial \beta} \right) = \lambda \left. \frac{\partial F_{00}^\mu}{\partial \alpha} \right|_{\lambda} = \hat{S}^\mu(\alpha,\beta),
\end{equation}
where we have noted that the combination of $\alpha$ and $\beta$ derivatives present in \eqref{jacx} corresponds to an $\alpha$ derivative at fixed $\lambda=\alpha \beta$, and we also have used the fact that since the left side is independent of $n^i$ and $t$, the right side also must be independent of $n^i$ and $t$.  Again, writing
\begin{equation}
\hat{S}^\mu(\alpha,\beta) = \hat{S}^\mu_0(\alpha) + \beta \hat{S}^\mu_1(\alpha,\beta) \, ,
\end{equation}
we find by the same type of argument as made several times above that $\hat{S}^\mu_0 = 0$. Thus, changing variables from $(\alpha, \beta)$ to $(\alpha, \lambda)$, we obtain
\begin{equation}\label{moocow}
\frac{\partial}{\partial \alpha} F^\mu_{00}(\alpha,\lambda/\alpha) = \frac{\hat{S}_1^\mu(\alpha,\lambda / \alpha)}{\alpha},
\end{equation}
where the derivative is at fixed $\lambda$.

We now claim that the general solution to \eqref{moocow} that is smooth in $\beta$ near $\beta=0$ at fixed $\alpha>0$ is
of the form
\begin{equation}
F_{00}^\mu = \check{S}^\mu(\alpha,\beta) + H^\mu (\lambda) \log \alpha
\label{cow}
\end{equation}
where $H^\mu$ is a smooth function of $\lambda$ (including at $\lambda = 0$).
To show this, we integrate \eqref{moocow} with initial conditions $f^\mu(\lambda) = F^\mu_{00}(\alpha_0,\lambda/\alpha_0)$ specified at some $\alpha_0>0$.  Since $F^\mu_{00}(\alpha_0,\beta)$ is known to be smooth in $\beta$ at fixed $\alpha_0>0$, it follows that $f^\mu$ is smooth.  For any $N,M$, we may write
\begin{equation}
\hat{S}_1^\mu(\alpha,\beta) = \sum_{n,m=0}^{N,M} a^\mu_{nm} \alpha^n \beta^m + \alpha^{N+1} \beta^{M+1} \hat{S}'^\mu(\alpha,\beta) \, .
\label{seriesS}
\end{equation}
We then replace $\beta$ by $\lambda/\alpha$, substitute into \eqref{moocow}, and perform the integration. We thereby obtain in a neighborhood of $\alpha=\beta=0$ the general solution of the form
\begin{equation}
F_{00}^\mu = f^\mu(\lambda) + \check{S}^\mu_{NM}(\alpha,\beta) + H^\mu_{NM}(\lambda) \log \alpha + R_{NM}(\alpha, \beta) \, .
\label{moo}
\end{equation}
Here, the first term arises from the initial conditions, the second and third terms arise from the explicit integration of the finite sum in \eqref{seriesS}, and the last term, $R_{NM}(\alpha, \beta)$, arises from the integration of the remainder term $\alpha^{N+1} \beta^{M+1} \hat{S'}^\mu(\alpha,\beta)$ in \eqref{seriesS}.
The last term can be seen to be $C^M$ in $(\alpha, \beta)$ in a neighborhood of $(0,0)$.  Since \eqref{moo} holds for all $N,M$, it follows that $F_{00}^\mu$ is of the form \eqref{cow}, as claimed. 

Thus, we have shown that imposition of the condition that the Jacobian 
$\partial x'^{\mu'}/\partial x^{\mu}$ be smooth in $(\alpha,\beta,n^i,t)$ implies that the transformation must
be of the form
\begin{equation}\label{coords}
{x'}^\mu = x^\mu + \lambda S^\mu(\alpha,\beta,n^i,t) + \lambda H^\mu (\lambda) \log r \, ,
\end{equation}
where $S^\mu$ is smooth in $(\alpha,\beta,n^i,t)$ including at $\alpha = \beta = 0$ and $H^\mu$ is smooth in $\lambda$ including at $\lambda=0$. However, if $H^\mu \neq 0$, then the transformation \eqref{coords} will {\em not}, in general, yield an allowed transformation, since the non-smoothness of the transformation in $r=\alpha$ will result in non-smoothness of $g'_{\mu' \nu'}$ in $(\alpha', \beta')$ except in very special cases, such as when $g_{ab}(\lambda)$ has a translational symmetry for all $\lambda$. Thus, in general, only the above transformations
with $H^\mu = 0$ are allowed. (This is the caveat to the claim of \cite{gralla-wald} mentioned in the second paragraph above.) On the other hand, if $H^\mu = 0$, i.e., if
\begin{equation}
{x'}^\mu = x^\mu + \lambda S^\mu(\alpha,\beta,n^i,t) \, ,
\label{allowed}
\end{equation}
then it is straightforward to check that $\alpha'$, $\beta'$, $n'^i$, and $t'$ are smooth functions of $(\alpha,\beta,n^i,t)$. Since this transformation is invertible near $\alpha=\beta=0$, it follows that $\alpha$, $\beta$, $n^i$, and $t$ can be expressed as smooth functions of $(\alpha',\beta',n'^i,t')$. Thus, $g_{\mu \nu}(\lambda, x)$ is a smooth function of the variables $(\alpha',\beta',n'^i,t')$. In addition, since the Jacobian approaches the identity as $(\alpha,\beta) \to (0,0)$, it follows that the inverse Jacobian $\partial x^\mu/\partial x'^{\mu'}$
is smooth in $(\alpha,\beta,n^i,t)$ and, hence, in $(\alpha',\beta',n'^i,t')$. Thus, 
under a coordinate transformation of the form \eqref{allowed}, $g'_{\mu' \nu'}(\lambda, x')$
is smooth in $(\alpha',\beta',n'^i,t')$, and all transformations of the form \eqref{allowed} are allowed.

Conversely, although we have not been able to prove this, we believe
that a necessary condition for an allowed transformation is that the inverse Jacobian $\partial x^{\mu}/\partial x'^{\mu'}$ be smooth in 
$(\alpha',\beta',n'^i,t')$, since we cannot see how non-smoothness of this quantity could be compensated
by non-smoothness of the original metric components $g_{\mu \nu}$ as functions of $(\alpha',\beta',n'^i,t')$
so as to produce a $g'_{\mu' \nu'}$ that is smooth in  $(\alpha',\beta',n'^i,t')$. Transformations for which
$\partial x^{\mu}/\partial x'^{\mu'}$ is smooth in 
$(\alpha',\beta',n'^i,t')$ must be of the form \eqref{coords} with the role of primed and unprimed variables reversed, i.e., 
\begin{equation}\label{coords2}
{x}^\mu = x'^\mu + \lambda S'^\mu(\alpha',\beta',n'^i,t') + \lambda H'^\mu(\lambda) \log r' \, ,
\end{equation}
Again, for an allowed transformation, we must, in general, have $H'^\mu =0$, in which case the inverse
transformation is of the form \eqref{allowed}. Thus, we believe that \eqref{allowed}---or, more generally, 
\eqref{allowed2} if we do not require the transformation to reduce to the identity at $\lambda=0$---is the necessary as well as sufficient form of an allowed transformation, for any nontrivial one-parameter-family $g_{ab}(\lambda)$.

For infinitesimal (i.e., gauge) transformations at first and second order in $\lambda$, equation \eqref{allowed} implies that the first and second-order gauge vectors $\xi_{(1)}^\mu$ and $\xi_{(2)}^\mu$ are smooth in $t,n^i$ at fixed $r$, and must take the form
\begin{align}
\xi_{(1)}^\mu & = F_{(1)}^\mu(n^i,t) + O(r) \label{xi1} \\
\xi_{(2)}^\mu & = \frac{F_{(2)}^\mu(n^i,t)}{r} + O(1) \label{xi2},
\end{align}
for smooth $ F_{(1)}$ and $F_{(2)}$.  Equation \eqref{xi1} is the form considered in \cite{gralla-wald}.  Equation \eqref{xi2} allows one to show that second order gauge transformations cannot affect the mass dipole (since a transformation of the given form cannot change the time-time component of the second-order metric perturbation at order $O(1/r^2)$), as claimed in \cite{gralla-wald}.

\end{document}